\documentclass[prd,aps,showpacs,superscriptaddress,floatfix,twocolumn]{revtex4-1}

\usepackage{comment}
\usepackage{hyperref}
\usepackage{mathrsfs}
\usepackage{amsmath}
\usepackage{amssymb}
\usepackage{epsfig}
\usepackage{epsf}
\usepackage{epsfig}
\usepackage{color}
\usepackage{ifpdf}
\usepackage{graphics}
\usepackage{longtable}
\usepackage{slashed}
\usepackage{float}

\newcommand{\ba}{\begin{eqnarray}}
\newcommand{\ea}{\end{eqnarray}}
\newcommand{\be}{\begin{equation}}
\newcommand{\ee}{\end{equation}}
\newcommand{\nn}{\nonumber}
\newcommand{\bi}{\begin{itemize}}
\newcommand{\ei}{\end{itemize}}

\begin{abstract}
Using the infinite-volume photon propagator, we developed a method which allows us to calculate electromagnetic corrections to stable hadron masses with only exponentially suppressed finite-volume effects.
The key idea is that the infinite volume hadronic
current-current correlation function with large time separation
between the two currents can be reconstructed by
its value at modest time separation, which can be evaluated in finite volume with only
exponentially suppressed errors.
This approach can be extended to other possible applications such as QED corrections to (semi-)leptonic decays and some rare decays.
\end{abstract}

\begin{document}

\title{
QED self energies from lattice QCD
without power-law finite-volume errors
}

\newcommand{\PKU}{
School of Physics, Peking University, Beijing 100871, China}
\newcommand{\CICQ}{
Collaborative Innovation Center of Quantum Matter, Beijing 100871, China}
\newcommand{\CHEP}{
Center for High Energy Physics, Peking University, Beijing 100871, China}
\newcommand{\LNPT}{
State Key Laboratory of Nuclear Physics and Technology, Peking University, Beijing 100871, China}

\newcommand{\UCONN}{
  Physics Department,
  University of Connecticut,
  Storrs, Connecticut 06269-3046,
  USA}
\newcommand{\RBRC}{
  RIKEN BNL Research Center,
  Brookhaven National Laboratory,
  Upton, New York 11973,
  USA}

\author{Xu Feng}
\affiliation{\PKU}
\affiliation{\CICQ}
\affiliation{\CHEP}
\affiliation{\LNPT}
\author{Luchang Jin}
\affiliation{\UCONN}
\affiliation{\RBRC}

\maketitle

\section{Introduction}

Electromagnetic and strong interactions are two fundamental interactions known to exist in nature. They are described by the first-principle theories of quantum electrodynamics (QED) and quantum chromodynamics (QCD), respectively. In some physical processes, QED and QCD are both present and both play indispensable roles. A typical example is the neutron proton mass difference, which is attributed to both electromagnetic and strong isospin-breaking effects. Although this mass difference is only 2.53 times the electron mass, it determines the neutron-proton abundance ratio in the early Universe, which is an important initial condition for Big Bang nucleosynthesis. This quantity attracts a lot of interest and has motivated a series of lattice QCD studies on the isospin breaking effects in hadron spectra~\cite{Duncan:1996xy,Duncan:1996be,Blum:2007cy,Blum:2010ym,Ishikawa:2012ix,Borsanyi:2014jba,Boyle:2017gzv}. 

Generally speaking, QED effects are small due to the suppression of a factor of the fine-structure constant $\alpha_{\mathrm{QED}}\approx1/137$. However, when the lattice QCD calculations reach the percent or sub-percent precision level, the QED correction becomes relevant. It plays a particularly important role in precision flavor physics, where lattice QCD calculations of the semi-leptonic decay form factors $f_+(0)$ and the leptonic decay constant ratio $f_K/f_\pi$ have reached the precision of $\lesssim 0.3\%$ \cite{Aoki:2016frl}. At this precision the isospin symmetry breakings cannot be neglected. Pioneering works~\cite{Carrasco:2015xwa,Lubicz:2016xro,Giusti:2017dwk} have been carried out to include QED corrections to leptonic decay rates.

The conventional approach to include QED in lattice QCD calculations is to introduce an infrared regulator for QED. One popular choice is QED$_\mathrm{L}$, first introduced in Ref.~\cite{Hayakawa:2008an}, which removes all the spatial zero modes of the photon field. There are also some other methods: QED$_\mathrm{TL}$~\cite{Borsanyi:2014jba}, massive photon~\cite{Endres:2015gda}, $C^*$ boundary condition~\cite{Lucini:2015hfa}. In general, by including the long-range electromagnetic interaction on a finite-volume lattice, all these treatments introduce power-law suppressed finite-volume errors. This is different from typical pure QCD lattice calculations where finite-volume errors are suppressed exponentially by the physical size of the lattice. Ref.~\cite{Davoudi:2018qpl} provides an up-to-date systematic analysis of the finite-volume errors for the hadron masses in the presence of QED corrections. 

Another approach to incorporate QED with QCD is to evaluate the QED part in infinite volume analytically and completely eliminate the power-law suppressed finite volume errors. Such an approach, called QED$_\infty$, has been used in the calculation of hadronic vacuum polarization (HVP) and the hadronic light-by-light (HLbL) contribution to muon $g-2$~\cite{Bernecker:2011gh,Asmussen:2016lse,Blum:2017cer,Blum:2018mom}.
This approach, when applied to QED corrections to stable hadron masses, does not completely remove the
power-law suppressed finite volume effects.
This is mostly because the hadron correlation functions, which one calculates
on the lattice
to extract hadron masses, are exponentially suppressed for large hadron source and sink separation.
Therefore, the finite volume error of the QED correction to the hadron correlation function
evaluated with QED$_\infty$, while its absolute size is still exponentially suppressed by the size
of the system,
is only power-law suppressed compared with the correction functions.
In this paper, we propose a method to solve this problem.
We show that the QED self-energy diagram can be calculated on a finite volume lattice
with only exponentially suppressed finite-volume effects.

\begin{figure}[H]
\begin{center}
\includegraphics[width=200pt,angle=0]{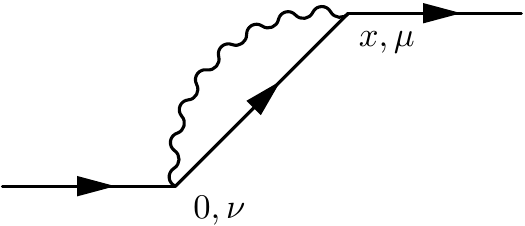}
\end{center}
\caption{Self-energy diagrams}
\label{fig:self_energy}
\end{figure}

\section{Master formula}

We first consider the self-energy calculation
in an infinite space-time volume.  For the case
of a stable hadronic state $N$, the self-energy
diagram shown in Fig.~\ref{fig:self_energy} can be calculated in
Euclidean space from the integral:
\be
\Delta M =
\mathcal{I}=
\frac{1}{2}\int d^4x\,\mathcal{H}_{\mu,\nu}(x) S^\gamma_{\mu,\nu}(x),
\label{eq:init}
\ee
where hadronic part $\mathcal{H}_{\mu,\nu}(x)=\mathcal{H}_{\mu,\nu}(t, \vec x)$ is given by
\be
\mathcal{H}_{\mu,\nu}(x)
=\frac{1}{2 M}\langle N(\vec 0)|T [J_\mu(x) J_\nu(0)]|N(\vec 0)\rangle,
\label{eq:def-H}
\ee
where $J_\mu=2e\, \bar{u}\gamma_\mu u/3 - e\, \bar{d}\gamma_\mu d/3 - e\, \bar{s}\gamma_\mu s/3$
is the hadronic current,
$|N(\vec p)\rangle$ indicates a hadronic state $N$ with the mass $M$ and spatial momentum $\vec p$,
and $S^\gamma_{\mu,\nu}$ is the photon propagator whose form is analytically known.
The states $|N(\vec p)\rangle$ obey the normalization convention
$\langle N(\vec p')|N(\vec p)\rangle = (2\pi)^3 2 E_{\vec p}\, \delta(\vec p' -\vec p)$.
The current operator $J_\mu(t, \vec x)$ is a standard Euclidean
Heisenberg-picture operator $J_\mu(t, \vec x) = e^{Ht}J_\mu(0, \vec x)\,e^{-Ht}$.
A possible short distance divergence of the integral can be removed by renormalizing the
quark mass.

If we examine an $L^3$ finite-volume system,
the main feature of the conventional methods such as QED$_{\mathrm{L}}$
is to design a finite-volume form for photon propagator, $S_{\mu,\nu}^{\gamma,L}$,
and calculate the hadronic correlation function
in a finite volume in the presence of finite-volume QED using $S_{\mu,\nu}^{\gamma,L}$.
Unfortunately, it results in power-law suppressed finite-volume effects in
the mass extracted from the finite-volume hadronic correlation function.
For the QED$_\infty$ approach,
one may begin with the infinite volume formula in Eq.~(\ref{eq:init}) to extract the QED self energy,
but then limit the range of the integral
and
replace $\mathcal{H}_{\mu,\nu}(x)$ by a finite volume version.
However, as we will explain later, the result still suffers from power-law finite-volume effects.

To completely solve the problem, we develop a method as follows.
We choose a time $t_s$ ($t_s \lesssim L$), that is sufficiently large
that the intermediate hadronic states between the two currents are dominated
by single hadron states
since all the other states (resonance states, multi-hadron states, etc)
are exponentially suppressed by $t_s$.
\ba
\mathcal{I}&=&\mathcal{I}^{(s)}+\mathcal{I}^{(l)},
\nn\\
\mathcal{I}^{(s)}&=&\frac{1}{2}\int_{-t_s}^{t_s}\!\! dt\int\! d^3\vec{x}\, \mathcal{H}_{\mu,\nu}(x) S^{\gamma}_{\mu,\nu}(x),
\nn\\
\mathcal{I}^{(l)}&=&\int_{t_s}^\infty\!\! dt\int\! d^3\vec{x}\, \mathcal{H}_{\mu,\nu}(x) S^{\gamma}_{\mu,\nu}(x).
\ea
We propose to approximate $\mathcal{I}^{(s)}$ and $\mathcal{I}^{(l)}$ using the lattice-QCD calculable expressions $\mathcal{I}^{(s,L)}$ and $\mathcal{I}^{(l,L)}$, which are defined as 
\ba
\label{eq:master_formula}
\mathcal{I}^{(s,L)}&=&\frac{1}{2}\int_{-t_s}^{t_s}\!\! dt\int_{-L/2}^{L/2}\!\! d^3\vec{x}\, \mathcal{H}^L_{\mu,\nu}(x) S^{\gamma}_{\mu,\nu}(x),
\nn\\
\mathcal{I}^{(l,L)}&=&\int_{-L/2}^{L/2}\!\! d^3\vec{x}\, \mathcal{H}^L_{\mu,\nu}(t_s,\vec{x}) L_{\mu,\nu}(t_s,\vec{x}),
\ea 
where $L_{\mu\nu}(t_s,\vec{x})$ is a QED weighting function, defined as:
\ba
&&L_{\mu,\nu}(t_s,\vec{x})
=
\int\!\!\frac{d^3p}{(2\pi)^3}
e^{i \vec{p}\cdot \vec{x}}
 \int_{t_s}^\infty\!\! dt\,
e^{-(E_{\vec{p}}-M)(t-t_s)}
\nn
\\ && \hspace{3cm} \times
 \int\! d^3\vec{x}'\,
e^{-i \vec{p}\cdot \vec{x}'}
S^\gamma_{\mu,\nu} (t,\vec{x}').
\label{eq:def-L}
\ea
Here the energy $E_{\vec{p}}$ is given by the dispersion relation $E_{\vec{p}}=\sqrt{M^2+\vec{p}^2}$. 
The integral in $L_{\mu,\nu}(t_s,\vec{x})$ can be
calculated in infinite volume (semi-)analytically.
In section~\ref{sect:evaluation_L}, detailed expressions 
for $L_{\mu,\nu}(t_s,\vec{x})$ are given 
for both Feynman- and Coulomb-gauge
photon propagators.

The finite-volume hadronic part $ \mathcal{H}^L_{\mu,\nu}(x) $ is defined
through finite-volume lattice correlators (assuming $t\ge 0$):
\ba
\!\!\!\!\!\!
\mathcal{H}^L_{\mu,\nu}(t, \vec{x})
=L^3\frac{\langle N(t+\Delta T) J_\mu(t,\vec x) J_\nu(0) \bar{N}(-\Delta T) \rangle_L}
{\langle N(t+\Delta T) \bar{N}(-\Delta T) \rangle_L},
\ea
where $\bar{N}(t)$/$N(t)$ is an interpolating operator which creates/annihilates
the zero momentum hadron state $N$ at time $t$,
$\Delta T$ is the separation between the source and current operators,
which only needs to be large enough to suppress the excited states effects.

We will demonstrate below the quantities
$\mathcal{I}^{(s,L)}$ and $\mathcal{I}^{(l,L)}$
defined in the master formula~(\ref{eq:master_formula})
only differ from $\mathcal{I}^{(s)}$ and $\mathcal{I}^{(l)}$
by exponentially suppressed finite-volume effects.

\section{Path to the master formula}

\subsection{Comparison between $\mathcal{I}^{(s)}$ and $\mathcal{I}^{(s,L)}$}

We adopt the conventional expectation (which can
be demonstrated in perturbation theory using the
Poisson summation formula \cite{Luscher:1985dn}) that for a theory such
as QCD with a mass gap a matrix element such as
$\mathcal{H}^L_{\mu,\nu}(t,\vec{x})$, evaluated in a
finite space-time volume $L^3\times T$ with periodic boundary
conditions, will differ from the corresponding
matrix element $\mathcal{H}_{\mu,\nu}(t,\vec{x})$ in infinite volume by terms that
are exponentially suppressed in the spatial and
temporal extents of the volume.
In addition,
the value of the infinite-volume $\mathcal{H}_{\mu,\nu}(t,\vec{x})$,
when $|\vec{x}|\gtrsim t$,
is exponentially suppressed in $|\vec{x}|$.

These considerations suggest that
the integral for $\mathcal{I}^{(s)}$
is dominated by the region inside the
finite-volume lattice and well
approximated by the finite-volume
integral $\mathcal{I}^{(s,L)}$.
We therefore conclude that $\mathcal{I}^{(s,L)}$ differs
from its infinite-volume version $\mathcal{I}^{(s)}$
by an exponentially suppressed finite-volume effect.

\subsection{Comparison between $\mathcal{I}^{(l)}$ and $\mathcal{I}^{(l,L)}$}
We remind the reader that the value of $\mathcal{H}_{\mu,\nu}(x)$
is not always exponentially suppressed at large $|x|$.
In fact, for large $|t|$, we shall have:
\be
\mathcal{H}_{\mu,\nu}(t,\vec{x})
\sim
e^{-M\left(\sqrt{t^2+\vec{x}^2}-t\right)}\sim e^{-M\frac{\vec{x}^2}{2t}}\sim O(1).
\ee
Therefore if we limit the range of the integral for $\mathcal{I}$ in Eq.~(\ref{eq:init}),
it will contain an $O(1/L)$
power-law finite volume effect even if
the infinite-volume photon propagator
$S_{\mu,\nu}^{\gamma}$ is used instead of
$S_{\mu,\nu}^{\gamma,L}$.
This is one of the reasons why the traditional
QED$_\infty$ method, which works for the cases of HVP and HLbL,
does not work for the QED self-energy diagram. 
As both ends of the photon propagator couple to the quark current,
one can only perform the intergral over a finite time window.
Even if we could create an infinite time-extent lattice and use the integral
\be
\int_{-\infty}^{\infty} dt\int_{-L/2}^{L/2}d^3\vec{x}\,{\mathcal{H}}^L_{\mu,\nu}(t,\vec{x})S_{\mu,\nu}^\gamma(t,\vec{x}),
\ee 
the result would still carry an $O(1/L^4)$ finite-volume effect, due to the fact that
 $\mathcal{H}^L_{\mu,\nu}(t,\vec{x})-\mathcal{H}_{\mu,\nu}(t,\vec{x})$ is not exponentially suppressed at large $|t|$.

Instead of using ${\mathcal{H}}^L_{\mu,\nu}(t,\vec{x})$ at large $|t|$ directly, we study the $t$-dependence of the infinite-volume $\mathcal{H}_{\mu,\nu}(t,\vec{x})$ for $|t|>t_s$.  By inserting a complete set of intermediate states, 
we can rewrite
$\mathcal{H}_{\mu,\nu}(t,\vec{x})$ as
\ba
\mathcal{H}_{\mu,\nu}(t,\vec{x})
&=&
\sum_n \int\!\!\frac{d^3\vec p}{(2\pi)^3}
\frac{
1}{2E_{n,\vec{p}}}
e^{i \vec{p}\cdot\vec{x}}
e^{-(E_{n,\vec{p}}-M)t}
\\
&&\hspace{-0.5cm}
\times
\frac{1}{2M}
\langle N(\vec{0})|J_\mu(0)|n(\vec{p})\rangle
\langle n(\vec{p})| J_\nu(0)|N(\vec 0)\rangle
.\nn
\ea
Without losing generality, positive $t$ is assumed in the above
equation.
In Euclidean space with large $t$,
the contribution from excited states si exponentially suppressed.
The following approximation,
where only the lowest energy states' contributions are kept,
is then valid for $t>t_s$:
\ba
\mathcal{H}_{\mu,\nu}(t,\vec{x})
&\approx&
\int\!\!\frac{d^3\vec p}{(2\pi)^3}
\frac{
1}{2E_{\vec{p}}}
e^{i \vec{p}\cdot\vec{x}}
e^{-(E_{\vec{p}}-M)t}
\label{eq:approx}
\\
&&\hspace{-0.5cm}
\times
\frac{1}{2M}
\langle N(\vec{0})|J_\mu(0)|N(\vec{p})\rangle
\langle N(\vec{p})| J_\nu(0)|N(\vec 0)\rangle
.\nn
\ea
where $ E_{\vec{p}}= \sqrt{M^2 + \vec{p}^2}$.
On one hand, Eq.~(\ref{eq:approx}) suggests that we can calculate $\mathcal{H}_{\mu,\nu}(t,\vec{x})$,
for large $t$,
via the matrix element $\langle M(\vec{p})|J_\mu(0)|M\rangle$. On the other hand, it indicates that
the Fourier transformation of $\mathcal{H}_{\mu,\nu}(t,\vec{x})$ at fixed $t=t_s$ gives the relevant matrix element:
\ba
&&
\int\! {d^3 \vec x}\,
\mathcal{H}_{\mu,\nu}(t_s,\vec{x})
e^{-i \vec{p}\cdot\vec{x}}
=
\frac{1}{2E_{\vec{p}}}
e^{-(E_{\vec{p}}-M)t_s}
\label{eq:H-ts}
\\&&
\hspace{1.5cm}
\times
\frac{1}{2M}
\langle N(\vec 0)|J_\mu(0)|N(\vec{p})\rangle
\langle N(\vec{p})| J_\nu(0)|N(\vec 0)\rangle
. \nn
\ea
Putting Eq.~(\ref{eq:H-ts}) into Eq.~(\ref{eq:approx}),
we are able to reconstruct the needed infinite volume
hadronic matrix element at large $t$
from its value at modest $t_s$:
\ba
&&\mathcal{H}_{\mu,\nu}(t,\vec{x}')
\approx
\int\! {d^3\vec x}\,
\mathcal{H}_{\mu,\nu}(t_s,\vec{x})
\label{eq:large-t} \\
&&\hspace{2cm}
\times\int\!\!\frac{d^3\vec p}{(2\pi)^3}
e^{i \vec{p}\cdot\vec{x}}
e^{-(E_{\vec{p}}-M)(t-t_s)}
e^{-i \vec{p}\cdot \vec{x}'}.
\nn
\ea
We will refer this relation, which
is the crucial step in the derivation, as the ``infinite volume reconstruction method''.
Here the $\approx$ symbol reminds us that the excited-state contributions in
$\mathcal{H}_{\mu,\nu}(t,\vec{x})$ and $\mathcal{H}_{\mu,\nu}(t_s,\vec{x})$ are exponentially suppressed and
have been neglected.

In the previous section, we have confirmed that $\mathcal{H}_{\mu,\nu}(t_s,\vec{x})$ is equal to
${\mathcal{H}}^L_{\mu,\nu}(t_s,\vec{x})$ up to some exponentially suppressed corrections
if $\vec{x}\in[-L/2,L/2]$, and is exponentially suppressed itself otherwise.
We therefore
conclude that $\mathcal{I}^{(l)}$ can be well approximated by $\mathcal{I}^{(l,L)}$ through
\ba
\mathcal{I}^{(l)}&=&\int_{t_s}^\infty dt \int d^3\vec{x}\,\mathcal{H}_{\mu,\nu}(t,\vec{x})S_{\mu,\nu}^\gamma(t,\vec{x})
\nn\\
&\approx&\int d^3\vec{x}\,\mathcal{H}_{\mu,\nu}(t_s,\vec{x})L_{\mu,\nu}(t_s,\vec{x})
\nn\\
&\approx&\int_{-L/2}^{L/2} d^3\vec{x}\,{\mathcal{H}}^L_{\mu,\nu}(t_s,\vec{x})L_{\mu,\nu}(t_s,\vec{x})
\nn\\
&=&\mathcal{I}^{(l,L)}
\ea
where the weighting function $L_{\mu,\nu}(t_s,\vec{x})$ has been given in Eq.~(\ref{eq:def-L})
and will be discussed in the following section.

\section{QED weighting function $L_{\mu,\nu}(t_s,\vec{x})$}
\label{sect:evaluation_L}
Detailed expressions
for the QED weighting function $L_{\mu,\nu}(t_s,\vec{x})$
defined in Eq.~(\ref{eq:def-L}) can be
evaluated
for Feynman- and Coulomb- gauge photon propagators:
\begin{itemize}
\item{Feynman gauge}
\be
S_{\mu,\nu}^\gamma(x)=\frac{\delta_{\mu,\nu}}{4\pi^2x^2}=\delta_{\mu,\nu}
\int\!\!\frac{d^4p}{(2\pi)^4}\,\frac{e^{ipx}}{p^2}.
\label{eq:feynman-prop}
\ee
\ba
\!\!\!
L_{\mu,\nu}(t_s,\vec{x})
&=&
\frac{\delta_{\mu,\nu}}{2\pi^2}
\int_0^\infty\!\!dp\,
\frac{\sin(p|\vec{x}|)}{2(p+E_p-M)|\vec{x}|} 
e^{-p t_s}.
\ea
%
%
\item{Coulomb gauge}
%
\ba
&&\quad
S^\gamma_{\mu,\nu}(t,\vec{x})
\label{eq:coulomb-prop}
\\
&&= \left\{
\begin{matrix}
\frac{1}{4 \pi |\vec{x}|} \delta(t) & \mu = \nu = 0\\
\int \frac{d^3\vec p}{(2\pi)^3} \frac{1}{2|\vec{p}|}
\left(\delta_{i,j} - \frac{p_i p_j}{\vec{p}^2} \right)
e^{-|\vec{p}|t + i \vec{p}\cdot\vec{x}}  & \mu =i, \nu = j\\
0 & \text{otherwise}
\end{matrix}
\right.
.
\nn
\ea
\ba
&&\quad L_{i,j}(t_s,\vec{x})\\
&&=
(\delta_{i,j}-\frac{x_i x_j}{\vec{x}^2})
\frac{1}{(2\pi)^2}
\int_0^\infty\!\!dp\,
\frac{\sin(p|\vec{x}|) }{2(p+E_p-M)|\vec{x}|} 
e^{-p t_s}
\nn\\
&&+
(\delta_{i,j}-3\frac{x_i x_j}{\vec{x}^2})
\frac{1}{(2\pi)^2}
\int_0^\infty\!\!dp\,
\frac{p|\vec{x}|\cos(p|\vec{x}|)-\sin(p|\vec{x}|) }{2(p+E_p-M)|\vec{x}|^3} 
e^{-p t_s}.
\nn
\ea

\end{itemize}
Only the spatial polarization components are needed
for the large time expression in Coulomb gauge.
All other components of $L$ are zero.

\section{Extended discussions}

Eq.~(\ref{eq:large-t}) tells us that the large time
hadronic matrix elements
$\mathcal{H}_{\mu,\nu}(t,\vec{x})$
can be determined using
$\mathcal{H}_{\mu,\nu}(t_s,\vec{x})$,
while $\mathcal{H}_{\mu,\nu}(t_s,\vec{x})$ can be
calculated using lattice.
Before reaching Eq.~(\ref{eq:large-t}), we explored other methods to determine
$\langle N(\vec 0)|J_\mu(0)|N(\vec 0)\rangle$. We recognized that by using the Coulomb-gauge photon propagator and assuming $|N(\vec 0)\rangle$ is a spin-0 charged particle, the corresponding matrix element can be determined easily. Here follows our discussion.

The infinite volume photon propagator in
Coulomb gauge is given in Eq.~(\ref{eq:coulomb-prop}).
This implies, for $\mathcal{I}^{(l)}$, only $S^\gamma_{i,j}$ is relevant.
For a spin-0 charged particle, we have
\ba
\langle N(\vec{p}_1)|J_\mu(0)|N(\vec{p}_2)\rangle = (p_1+p_2)_\mu F(q^2),
\ea
where the matrix element is expressed in terms of the form factor $F(q^2)$ with $q=p_1-p_2$.
If the initial or the final state has zero momentum, as is the case in Eq.~(\ref{eq:approx}), we have
\ba
\langle N(\vec{0})|J_i(0)|N(\vec p)\rangle = p_i F(p^2).
\ea
Therefore,
we can obtain that $\mathcal{I}^{(l)} = 0$
simply because of the Coulomb-gauge condition.
Thus
\ba
\!\!\!
\mathcal{I} &\approx& \mathcal{I}^{(s,L)} = 
\frac{1}{2}\int_{-t_s}^{t_s}\!\! dt \int_{-L/2}^{L/2}\!\! d^3\vec x\,
\mathcal{H}^L_{\mu,\nu}(t,\vec{x}) S^\gamma_{\mu,\nu}(t,\vec{x})
\ea
for a spin-0 charged particle and a Coulomb-gauge photon propagator,
and all the finite volume errors are exponentially suppressed by the lattice size, $L$,
or the integration range in the time direction, $t_s$.
Note that $t_s\lesssim L$ is required for the above statement to be valid.

\section{Conclusion}

We have demonstrated that the QED self-energy
for a stable hadron can be calculated on
a finite volume lattice with only
exponentially suppressed finite volume effects.
The power-law finite volume effects,
which are common in QCD+QED calculations,
are completely eliminated.
This is achieved with the following three ideas:
\begin{enumerate}
\item QED$_\infty$:
We start with an integral $\mathcal{I}$, where
the QED part in the integrand can
be calculated in infinite volume analytically,
and the hadronic part is purely a QCD matrix element,
and enjoys an exponential suppressed long distance behavior
because of the mass gap,
as is familiar from pure QCD lattice calculations;
\item Window method:
We introduce a cut in the time extent of the integral, $t_s$,
to separate the integral into the short-distance part,
which can be calculated within finite volume directly,
and the remaining long-distance part.
\item Infinite volume reconstruction method:
We use the fact that the long-distance
hadronic function is dominated by the lowest isolated pole
(the hadron whose QED mass shift is under study)
in the spectral representation
to express the infinite-volume hadronic function at large $t$
in terms of its value at modest $t_s$,
which can be evaluated in finite volume.
\end{enumerate}

The first idea, QED$_\infty$, has already been employed in
some QED+QCD calculations,
e.g. HVP \cite{Bernecker:2011gh},
HLbL \cite{Blum:2017cer,Asmussen:2016lse} and
the QED correction to HVP \cite{Blum:2018mom}.
For these calculations, this idea by itself is able to
remove all the power-law suppressed finite-volume errors.
The second idea used in this work, the window method, is relatively new.
The name of the method comes from Ref. \cite{Blum:2018mom},
where the integrand is also divided into parts,
and different treatments are applied to different parts.
The third idea, the infinite volume reconstruction method,
combined with the window method, is the essential
part of our framework.
It should be emphasized that, it is the \emph{infinite-volume}
hadronic function, $\mathcal{H}_{\mu,\nu}(t,\vec{x})$, at large $t$, being expressed in terms
of $\mathcal{H}_{\mu,\nu}(t_s,\vec{x})$ at modest $t_s$, which helps eliminate
the power-law finite-volume errors.

In additional to QED self-energy,
the framework developed here can also be applied to
other QED+QCD problems.
One example is the QED corrections to (semi-)leptonic decays,
which can be used to determine some important CKM matrix elements
like $V_{ud}$ and $V_{us}$~\cite{Carrasco:2015xwa,Lubicz:2016xro,Giusti:2017dwk}.
Another example is the rare kaon decays~\cite{Christ:2015aha,Christ:2016eae,Christ:2016mmq,Bai:2017fkh,Bai:2018hqu},
where the light electron propagator
can be treated in a similar way as the photon discussed in this paper
to reduce the finite-volume error.

\acknowledgements

We  would  like  to  thank  our  RBC  and  UKQCD
collaborators  for  helpful  discussions  and  support.

\appendix

\bibliographystyle{apsrev4-1}
\bibliography{ref.bib}

\end{document}